\documentclass[aps,prd,preprintnumbers,nofootinbib,twocolumn]{revtex4}
\usepackage{bm}
\usepackage{latexsym}
\usepackage{dcolumn}
\usepackage{amsmath,amsfonts,amssymb}
\usepackage{graphicx,epsfig}
\usepackage{psfrag}
\usepackage{amsthm}
\usepackage{color}

\def\be {\begin{equation}}
\def\ee {\end{equation}}
\def\bea {\begin{eqnarray}}
\def\eea {\end{eqnarray}}
\def\bc {\begin{center}}
\def\ec {\end{center}}
\def\bfg {\begin{figure}}
\def\efg {\end{figure}}
\def\bi {\begin{itemize}}
\def\ei {\end{itemize}}

\def\no {\noindent}

\def\vs {\vspace}

\newcommand{\bdm}{\begin{displaymath}}
\newcommand{\edm}{\end{displaymath}}

\begin{document}

\title{There is no coincidence after all!}

\author{Saurya Das} \email[email: ]{saurya.das@uleth.ca}

\vs{0.3cm}

\affiliation{Theoretical Physics Group,
Department of Physics and Astronomy,
University of Lethbridge, 4401 University Drive,
Lethbridge, Alberta T1K 3M4, Canada \\}

\begin{abstract}
We show that if Dark Matter is made up of light bosons, they form a Bose-Einstein condensate in the early Universe. This in turn
naturally induces a Dark Energy of approximately equal density
and exerting negative pressure.
This explains the so-called coincidence problem.

\vspace{0.3cm}
\noindent
{\bf This essay received an Honorable Mention in the 2020 Gravity Research Foundation Essay Competition}\\
{}\\
{}\\

\end{abstract}

\maketitle

We ask a simple question: why are the Dark Matter (DM) and Dark Energy (DE)
energy densities approximately equal, i.e. $\rho_{DM}\approx \rho_{DE}$? Since their ratio should go as $\rho_{DM}/\rho_{DE} \propto 1/a^3$ (where $a$ is the scale factor), this seems to suggest, inexplicably, that we are in a special epoch with a special value of $a$ for which the ratio is approximately equal to unity. While nothing in principle prevents this from happening, at the very least, this goes against the widely held and supported view that our place and time in the universe are not special, and
definitely warrants a better understanding.

Noting that DM and DE exert positive and negative pressures (potentials) respectively, we now ask a related question:
is there a natural mechanism by
which a positive potential somehow induces an equal and opposite negative potential or vice-versa? The answer is remarkably in the positive.
The Schr\"odinger equation with a classical potential $V$
can be written as Newton's law with an additional `quantum potential'
$V_Q$, i.e.
\begin{eqnarray}
m\, \ddot{\vec r} = -\vec\nabla (V+V_Q)~.
\label{nl1}
\end{eqnarray}
where for the wavefunction for the system
$\psi={\cal R}\,e^{iS}$, $({\cal R},$ $S=\mbox{Real functions})$,
$V_Q = -(\hbar^2/2m)\,(\nabla^2 {\cal R})/{\cal R}$.
All physical predictions of the above `quantum corrected Newton's equation'
is identical to those of the Schr\"odinger equation with just the potential $V$ \cite{bohm}.
Furthermore, for stationary states,
the quantum potential is exactly equal and opposite to the starting classical potential, i.e. $V_Q=-V$! The dynamics from Eq.(\ref{nl1}) is thus generated by deviations for stationarity or in general by superposed quantum states.
Motivated by this, it seems reasonable to expect that if the
quantum corrections to the classical Friedmann equation can be written as additional
density and pressure terms,
then at least for certain states of the DM fluid,
a `quantum fluid' with negative pressure may result.

Fortunately, such a quantum corrected Friedmann/Raychaudhuri equation
has been derived recently \cite{sd,db1,db2,db3}
\begin{eqnarray}
\frac{\ddot a}{a}
= - \frac{4\pi G}{3}\rho_{crit}
+ \frac{\hbar^2}{3m^2}\, h^{\mu\nu}
\left( \frac{\Box {\cal R}}{{\cal R}}\right)_{;\mu;\nu}~,
\label{ho4}
\end{eqnarray}
where $h_{\mu\nu} = g_{\mu\nu} - u_\mu u_\nu$ is the induced metric in terms of the `velocity field' of the DM, $u_\mu=\partial_\mu\,S/m$.
The first term on the RHS is the usual DM term, with
$\rho_{DM} \approx \rho_{crit}=3H_0^2/8\pi G$,
the critical density (where $H_0$ is the Hubble parameter).
To check if the last term can be interpreted as DE, we first construct the DM wavefunction $\psi$.
Consider a DM particle of mass $m$ on the surface of a
sphere of radius $r$ in the comoving frame. Only the mass inside the sphere will contribute to the gravitational potential to which it is subject.
Now of course the density inside the sphere will decrease (increase) for the
expanding (contracting) branch of the solution to the classical
Friedmann equations, but for times scale $\tau \ll H_0^{-1}$ ,
this variation is small and
can be ignored, and the potential is that of a
simple harmonic oscillator with frequency $\omega=H_0/\sqrt{2}$.
Note that this covers a significant fraction
of the age of the universe.
The wavefunction is then a superposition of
$\psi_n \propto \exp(-m\,\omega\, r^2/2)\,H_n(\sqrt{m\omega}\,r)$,
where the $H_n$\,s are the Hermite polynomials.
But what guarantees that the total $V_Q$ induced by the aggregate of DM
particles will coherently add up and not cancel each other?
This is easily solved if one assumes that the particles are light bosons.
In fact, it was shown in \cite{db1} that
for $m \lesssim \,6\,eV/c^2$, the temperature of the Universe at any epoch
$T (a)\leq T_c (a)$, the critical temperature
below which the bosons form a Bose-Einstein condensate (BEC) at that epoch. Therefore it follows that the
DM would form a BEC in the early Universe, and
the vast majority of the constituent bosons
will be in their ground states.
A number of authors in the past have considered BEC DM
(see e.g. \cite{hu}, \cite{sikivie} and references in
\cite{db2}).
However, to the best of our knowledge, the emergence of
DE via its quantum potential was not considered.
This is described by the macroscopic wavefunction spanning across cosmological length scales,
$\psi = R(a)\,e^{-m\,\omega\,r^2/2}={\cal R}~,$ where
$R(a)=R_0/[a(t)]^{3/2}$, to account for the
$\rho_{DM}\propto 1/a^3$ decay of DM density.
The above wavefunction, when substituted in Eq.(\ref{ho4}),
yields
\begin{eqnarray}
\frac{\ddot a}{a} &&= -\frac{4\pi G \rho_{crit}}{3}
+ \frac{H_0^2}{2}
\\
&& =-\frac{4\pi G}{3}
\left( \rho_{DM} + \rho_{DE} + 3\,p_{DE} \right)
\nonumber \\
&&= 0~ \label{fe4}
\end{eqnarray}
where $\rho_{DM} = \rho_{crit},~\rho_{DE}= 0.5\,\rho_{crit}$
and $p_{DE} = -\rho_{DE}$.
In other words, the density of the induced DE is (approximately) the same as that of DM, and it exerts negative pressure.
In real life, $\rho_{DM}$ would start decaying as $1/a^3$ almost as soon
as the equality (\ref{fe4}) is obeyed.
This and the fact that a small fraction of the bosons are in the
excited states should explain the precise ratio $\rho_{DM}/\rho_{DE}$
that has been observed.
Note that the mass $m$ cancels out and
our final result does not depend on it as long as it is light enough to form a BEC.
That said, the behaviour of DM BEC in galaxies and other regions of strong gravity may impose constraints on $m$, and the detection of such particles would be a vindication of our model.
These apart, our simple model shows that under practically a single assumption, namely that DM is made up of light bosons, the coincidence problem has a natural resolution.
This is simply because quantum mechanics
induces DE from DM of an equal density and negative pressure.


\noindent
{\bf Acknowledgment}

\no
This essay is dedicated to the memory of Professor R. K. Bhaduri,
with whom the main results were jointly derived.
This work was supported by the Natural Sciences and Engineering
Research Council of Canada.
%



\end{document}